# Automated visual inspection of silicon detectors in CMS experiment

Dr. Nupur Giri
*Dept. of Computer Engineering*
*Vivekanand Institute of Technology*
Mumbai, India
nupur.giri@ves.ac.in

Prof. Shashi Dugad
*Dept. of High Energy Physics*
*Tata Institute of Fundamental Research*
Colaba, Mumbai, India
shashi@tifr.res.in

Amit Chhabria
*B.E. Computer Engineering*
*Vivekanand Institute of Technology*
Mumbai, India
2018.amit.chhabria@ves.ac.in

Rashmi Manwani
*B.E. Computer Engineering*
*Vivekanand Institute of Technology*
Mumbai, India
2018.rashmi.manwani@ves.ac.in

Priyanka Asrani
*B.E. Computer Engineering*
*Vivekanand Institute of Technology*
Mumbai, India
2018.priyanka.asrani@ves.ac.in

*Abstract*—In the CMS experiment at CERN, Geneva, a large number of HGCAL sensor modules are fabricated in advanced laboratories around the world. Each sensor module contains about 700 checkpoints for visual inspection thus making it almost impossible to carry out such inspection manually. As artificial intelligence is more and more widely used in manufacturing, traditional detection technologies are gradually being intelligent. In order to more accurately evaluate the checkpoints, we propose to use deep learning-based object detection techniques to detect manufacturing defects in testing large numbers of modules automatically.

*Index Terms*—Automated visual inspection, CMS, Object Detection, CMM, Yolov4-tiny, VGG16.

## I. INTRODUCTION

CERN's Large Hadron Collider (LHC)[1] is the most powerful and largest particle accelerator. It accelerates protons to nearly the speed of light in clockwise and anti-clockwise directions. It collides them at four different locations situated around its ring.It was designed to explore the miscellaneous concepts of proton particle collision among them. At the center of the CMS detector, a huge bore superconducting solenoid consisting of a high magnetic field, surrounds the silicon component and practical tracker. It also encircles a brass-scintillator sample hadron calorimeter and a lead-tungstate gleaming crystal electromagnetic calorimeter. LHC boosts particles nearly to the speed of light, so the proton particles can collide with each other. Collisions of these particles produce immense amounts of energy and signals are then passed to detectors for digitalization. The digitization is done by the electronic components which are mounted on the silicon detector. So the signal is to be carried from the detector to the electronic component on it. That is carried through the bonds which are fabricated on the sensor modules.

To evaluate these sensors various Deep Learning based object detection techniques are built, it tells us about bonds present in the sensors are properly intact to each other and thus implements automated visual inspection.

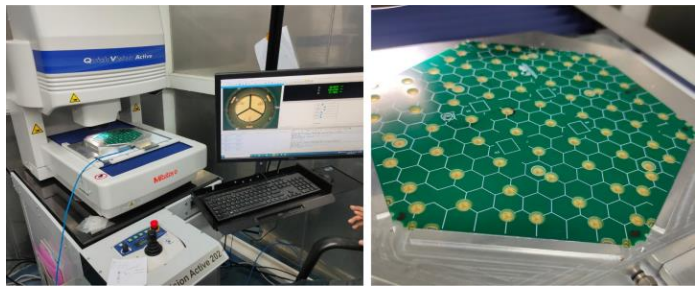

Fig. 1: Sensor Board

When comparing manual and automatic visual inspection (AVI) [2,3], AVI saves time for human inspectors. Manual inspection is time-consuming and expensive process which results in high scrap rates and a lack of quality assurance. Also, the manual inspection is not practical as production rates in high-tech industries are very high. AVI is a contactless inspection approach that prevents the inspection process from causing mechanical harm. The primary purpose of visual-based techniques [4] is to comprehend the world in both natural and man-made forms. The process of identifying images in the latter is mostly an attempt to find a mathematical/logical connection between the input images and environmental representations.

## II. LITERATURE REVIEW

In the regards of Rau, H., Wu, CH. Automatic optical inspection for detecting defects on printed circuit board inner layers. Int J Adv Manuf Technol as seen in [5] discusses Automatic visual evaluation for identifying faults for PCB inner layers. The authors used five different stages: The first step was to reconstruct the reference image, which was followed by inspection image normalisation. Image subtraction and defect separation are the next steps, followed by defect classification and they also proposed two-step image processing inspection. In the expansion and contraction procedure, an image subtraction operation is used to compute the difference between predicted images and actual images, with appropriate deviation according to design standards. Shorts, missing conductors, pinholes, excess copper, mouse bites, spurs, Opens and missing holes are among the eight types of defects that can be detected by an inspection system.

With respect to automated visual inspection, the authors Qiwu Luo, Xiaoxin Fang, Jiaojiao Su, Jian Zhou, Bingxing Zhou, Chunhua Yang, Li Liu, Weihua Gui as seen in the [6] explains that quality inspection of materials, defect classification is an essential stage after defect detection. So, the report investigates about 140 investigations on three distinct flat steel products: hot-rolled and cold-rolled steel strips, con-casting slabs, provides a review of established and developing computer vision based automated techniques for defect detection. The steps followed for defect detection and classification are image acquisition through the camera, light source and lightning methods, image preprocessing like image denoising, image segmentation, and enhancement, feature extraction based on transforming, and deep learning approaches like grayscale methods, texture-based methods, deep learning techniques, etc. The next step is feature selection which involves feature evolution, dimension reduction, and feature optimization, and the last step is defect classification using supervised and unsupervised algorithms. In the image acquisition procedure, they had taken acceptable measures to overcome the problem of low image quality. Image preprocessing includes applying the latest techniques to enhance image quality. For feature extraction, a combination of various methods has been implemented to successfully extract the features. Feature selection is an essential part of the process to remove redundant and irrelevant features of the image. In terms of defect classification, priority was on creatively combining various studied features and using the taxonomic potentials of the various machine learning algorithms to develop adaptive highly precise defect classification schemes, generalisation, and robustness.

In regards to the process of Automated Visual Inspection authors Roland T. Chin, Charles A. Harlow discusses Automated visual inspection image processing techniques for quality control and production line automation as seen in [7]. A detailed explanation regarding gradual improvement by applying feature-based approaches is given. The fast advancement of manufacturing technology has resulted in increasingly complex, smaller chips with more sophisticated architectures. The paper also discussed development and usage of super resolution imaging techniques for inspecting small regions and intricate chip components.

## III. RELATED WORKS

### A. Object Detection

Object detection [8] is a process that identifies the objects in an image or video it also also locates the position of the objects. Specifically, object detection creates surrounding rectangular boxes around detected objects, allowing us to determine where these objects are in a given frame. Object detection is inextricably linked to other similar computer vision techniques such as image identification and image bifurcation in that it allows us in comprehending and analyzing scenes in images or video.

Object detection [9,10] is broadly classified into two approaches i.e. ML-based and DL-based. In more conventional ML-based approaches, computer vision algorithms are used to detect groupings of pixels that may belong to an object by examining different aspects of an image, such as the color histogram or edges.

These extracted features then act as an input for a regression model, which finds the object coordinate and annotates it. On the other hand DL-based approaches utilizes convolutional neural networks (CNNs) to carry out unsupervised object detection, which inturn eradicates the requirement for distinct feature definition and extraction.

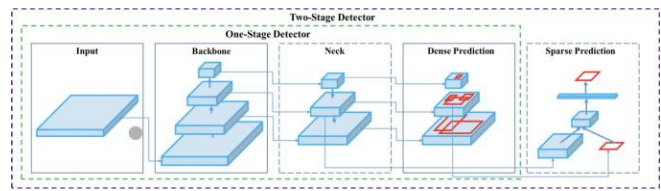

Fig. 2: Architecture of object detection algorithms

### B. VGG16

VGG16 [11] is a commonly used Convolutional Neural Network (CNN) architecture that was developed for ImageNet, a large visual database project used in the development of object detection software. Karen Simonyan and Andrew Zisserman of the University of Oxford developed and released the VGG16 Architecture in their publication "Very Deep Convolutional Networks for Large-Scale Image Recognition" in 2014.

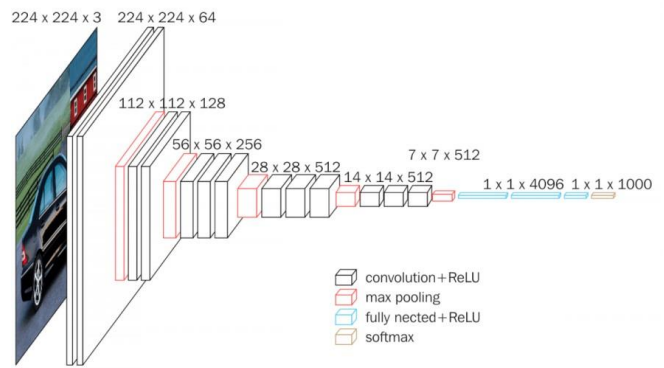

Fig. 3: Architecture of VGG16 Algorithm

VGG16 is noteworthy because inspite of having a massive amount of hyper-parameters, they concentrated on having 3x3 filter convolution layers with a stride of 1 and always utilized padding and maxpool layer having a 2x2 filter with a stride of 2. In the architecture the maxpool and convolution layers are placed in the same order throughout the system. The output is given at the softmax layer, which is present at the end of two fully connected layers. As a matter of fact 16 in VGG16 stands for 16 layers having weights.

*C. YOLO*

You Only Look Once is a one-stage detector proposed by Redmond et al. Object detection techniques before YOLO used remodeled classifiers to execute detection, while YOLO tends to use neural network convolution layers that perform and give the predictions in a rectangular box and also calculate the probability of classes all at once. YOLO [15] has the inherent benefit of speed when compared to real-time object detectors, in addition to greater predictive performance and a sharper Intersection over Union (IOU) in bounding boxes.Other algorithms utilize the Region Proposal Network to predict expected region of interest. Recognition is done independently in these regions. Whereas YOLO performs all of its predictions using a single fully connected layer. As a result, Region Proposal Networks algorithms use a large number of iterations for predicting an image, whereas YOLO takes a single iteration for the same image to detect the bounding boxes around the object.

The YOLO method is to divide the image into N numbers of grids, where each grid is of the same dimensional region of SxS. Each of the N grids is responsible for recognising and identifying the object it contains. In turn, these grids calculate bounding box coordinates in reference to cell coordinates. It also predicts the annotation and possibility of the object being present in the grid. This method remarkably minimizes computation because both recognition as well as detection are managed by grids from the image. However, because many grids detect the same object but have different bounding box predictions, it creates a lot of duplicate predictions. To address this issue, YOLO employs Non-Maximal Suppression. In Non-Maximal Suppression, YOLO removes all anchor boxes with lower threshold probability values.The algorithm accomplishes this by first looking at the probability scores and then selecting highest probability among all the other scores for each decision. The bounding boxes with the high probability scores and highest Intersection over Union are suppressed first for getting more accurate results, because high IOU and probability scores suggest pointing to the same object. This process is continued until the required bounding boxes are obtained.

Yolo [13,14] has several variants such as YOLOv2 is an improved version of YOLOv1. By implementing batch normalization, YOLOv2 improves the network's mean Average Precision (mAP) by as much as 2 percent. To increase accuracy of tiny objects, YOLOv3 [15] introduced an probability score to bounding box prediction, added connections to the backbone network layers, and produced predictions at three different degrees of granularity.

*D. YOLOv4-tiny*

YOLOv4-tiny algorithm [16] is an object detection algorithm based on the YOLOv4 model. Yolov4 tiny was proposed as a way to improve network efficiency by simplifying the network structure and reducing parameters. It is used for object detection, and it works well in real-time.

YOLOv4-tiny is a compact version of YOLOv4 that is intended for use on machines with limited computing power. It enables us to deploy applications on mobile and embedded devices due to the reduced parameters. The model weights are approximately of size 16 MB, which allows it to train 350 images in an hour on a Tesla P100 GPU. On the Tesla P100, the inference speed of YOLOv4-tiny is 3 ms, making it one of the fastest object detection models available [17].

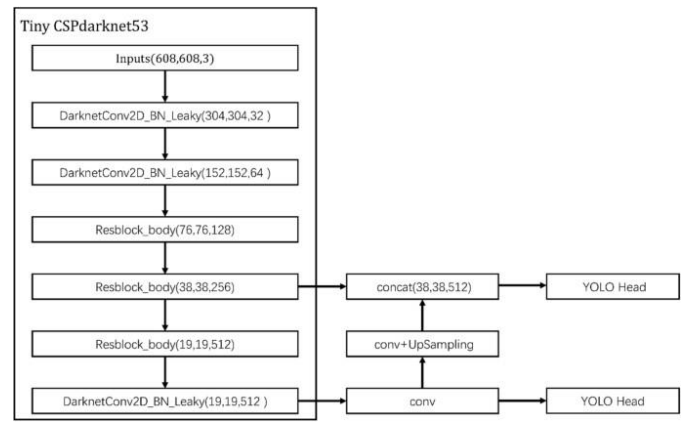

Fig. 4: Yolov4-tiny Architecture

To achieve these fast speeds, YOLOv4-tiny [18] has different changes than the original YOLOv4 network. The main difference between YOLOv4 tiny and YOLOv4 is the network size reduction. It has YOLO heads, compact to three in YOLOv4, resulting in fewer YOLO layers and fewer anchor boxes for prediction. It uses 29 pre-trained convolutional layers, whereas YOLOv4 uses 137 pre-trained convolutional layers, resulting in a reduction in the number of convolutional layers in the CSP backbone. Using the Yolov4 model and the Darknet framework and train the model.

FPS (Frames per Second) in YOLOv4-tiny is roughly 8 times that of YOLOv4, indicating that inference time is reduced and it is roughly 2/3 as fast on MS COCO (a very hard dataset). Even less performance degradation on small custom detection tasks that are more tractable. On the RTX 2080Ti, the YOLOv4-tiny model achieves 22.0 percent AP (42.0 percent AP50) at 443 FPS, while the YOLOv4-tiny achieves 1774 FPS using TensorRT, batch size = 4, and FP16-precision.

YOLOv4-tiny can be used for much faster training and detection. Using a smaller model, such as YOLOv4-tiny, is the most effective way to reduce your model's inference time. It's great for making object detector mobile apps because yolov3, yolov4 or other heavy models cannot be used for mobile apps.

## IV. METHODOLOGY

### A. Data Collection

The CMS Experiment apparatus has provided the data. The images include silicon detectors with wire bonds configured in a group of threes. These images have been created with the help of Coordinate Measuring Machine (CMM) which measures the center point, orientation, and the location of detectors w.r.t reference points marked on the mounting board. The CMM gives the geometry of silicon detectors with an accuracy of about 15 μm. The main aim of the project is to determine whether the bonds are totally intact as well as parallel with each other, and at the same time to check whether the sensors present on the PCB board are free of calibration and glue for quality control.

### B. Data Generation

The dataset comprises 169 images for step 1 and 346 images for step 2 having pixel resolution of 1280 x 1024. To proceed further, the images must be labeled with the corresponding class. LabelImg of the images were performed by the annotation tool which depicts the area of interest to the object detection algorithms. While annotating the images, bounding box must be drawn and assign the class to that specific area. Multi class classification approach for annotation and object detection has been used.

For step 1, classes were:
- 12_4_8_c - Orientation of bonds is 12_4_8 and calibrated
- 12_4_8_nc - Orientation of bonds is 12 4 8 and not calibrated
- 2_6_10_c - Orientation of bonds is 2_6_10 and calibrated
- 2_6_10_nc - Orientation of bonds is 2 6 10 and not calibrated

For step 2, the classes were:
- three - three intact and parallel wire bonds
- not three - broken or not-parallel wire bonds
- glue - images contains glue drop
- no_bonds - no bonds present

Once the annotations are performed, they can be exported in txt file. Txt files are used for YOLO algorithms. These files consist of the class name, centre x, centre y, width and height of the bounding box. After that, train and test set is created by slicing the dataset in the ratio 8:2.

### C. Data Modelling

For the quality assurance of sensors, two-step process has been followed. The first step is a classification of orientation and calibration using the VGG16 algorithm and the second step is the detection of glue and broken bonds using the YOLOv4-tiny algorithm.

*1) VGG16:* VGG16 is a CNN model widely used for transfer learning which helps for training and building classification models. The VGG16 model contains pretrained weights and it uses those pre-trained weights to perform transfer learning and train other models with respect to their dataset. Hence this model has been used to classify the images and check whether the images have a calibration spot on them at the same time, it will also check the orientation of the images. As there are four classes(12_4_8_c, 12_4_8_nc, 2_6_10_c, 2_6_10_nc), we have to create four folders respectively and classify our dataset into these folders. The name of the folder acts as a label for the images. 169 images were used for training the model.

*2) YOLOv4-tiny:* YOLOv4 tiny is roughly 8X as fast at inference time as YOLOv4. In the YOLOv4-tiny network, size is reduced and convolutional layers in the CSP backbone are compressed. For YOLOv4-tiny yolov4-tiny-custom.cfg was used with batch size 64 and subdivisions 16 for 8000 epochs. The weight file used to train the Yolov4 tiny is yolov4-tiny.conv.29. 346 images were used for training which contains 4 classes such as three, not_three, glue, and no_bonds which will tell the actual state of the sensor for the quality control i.e the bonds present on the PCB are broken, intact or missing.

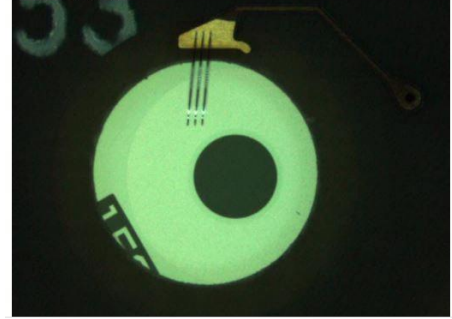

(a)

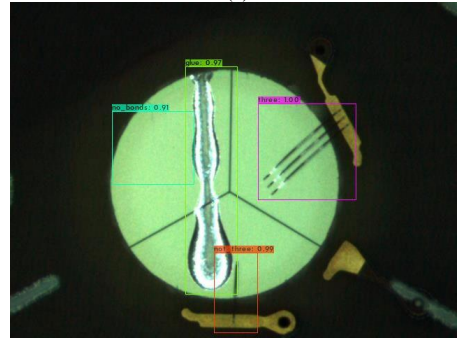

(b)

Fig. 5: Results

## V. RESULTS

Figure 5 illustrates the final classification and the associated decision boundary for the wire bonds which gives us a gist of quality assurance of the senors. From these results, images that were not included in the initial training or testing set can be observed and automatically classified as three, not three, glue, and no_bonds. It will also let us know about the orientation and calibration type of image using this object detection model.

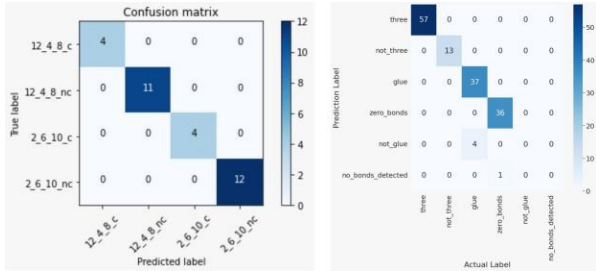

Fig. 6: Confusion Matrix - (a) VGG16 (b) TinyYOLOv4

The table below shows the final outcome of training, which includes Accuracy, Mean Average Precision (mAP), and False Positive rate for various methods. In addition to accuracy and mean average precision False Positive rate was also calculated which denotes bad quality bonds detected as good quality because it is an important evaluation metric in our case.

The performance of a model on a testing dataset for which the true values are known is described by a confusion matrix. 31 images were used for testing the orientation and calibration type of image, 35 images were used to check whether the wired bonds are intact and parallel to each other. In the confusion matrix, the vertical axis reports the predicted labels, and the horizontal axis reports actual labels. Figure 6(a) describes the confusion matrix for VGG16 and Figure 6(b) describes the confusion matrix for Tiny YOLOv4. These trained models discover the defect with an accuracy of 100% and 96.6%.

TABLE I: Results

| Algorithm | Accuracy | False Positive | mAP |
|---|---|---|---|
| VGG16 | 100 | 0 | 100 |
| Yolov4-tiny | 96.6 | glue-9.7, no bonds-2.7 | 97.39 |

## CONCLUSION

This project lays the groundwork for building a system that can automatically detect defects in the wire bond and ensure quality of the system. Quality control is an important part of production and research on defect detection has great practical significance to ensure product quality. We've provided a comprehensive literature survey on automated visual inspection using object detection techniques.

For this project, a dataset was obtained from the detector. The labeled data was then used as training and testing data using the object detection algorithm. The results from the system can be used to automate the process of wire bond defect detection.

Limitations of proposed work include situations when there is too much noise in the image or type of detector is changed. Future study will attempt to overcome these shortcomings and attain accuracy that is equivalent to or nearly equal to ground truth.